\documentclass[prx,twocolumn,groupedaddress]{revtex4-1}
\usepackage{graphicx}
\usepackage{dcolumn}
\usepackage{bm}

\usepackage{amsmath}
\usepackage{graphicx}
\usepackage{color}
\usepackage{wrapfig}
\usepackage{epstopdf}

\begin{document}

\title{Approach to tune near- and far-field properties of hybrid dimer nanoantennas via laser melting at the nanoscale}

\author{Yali Sun,$^{1,2}$ Stanislav Kolodny,$^{1}$ Dmitry Zuev,$^{1}$ Lirong Huang,$^{2}$ Alexander Krasnok,$^{1}$ and Pavel Belov$^{1}$}
\address{
$^{1}$ITMO University, St.~Petersburg 197101, Russia\\
$^{2}$Wuhan National Laboratory for Optoelectronics, Huazhong University of Science and Technology, Wuhan, Hubei, 430074, China}

\begin{abstract}
Asymmetric metal-dielectric nanostructures are demonstrated superior optical properties arise from the combination of strong enhancement of their near fields and controllable scattering characteristics which originate from plasmonic and high-index dielectric components. Here, being inspired by the recent experimental work [Dmitry~Zuev, \textit{et al.}, Adv. Mater. \textbf{28}, 3087 (2016)] on a new technique for fabrication of asymmetric hybrid nanoparticles via femtosecond laser melting at the nanoscale, we suggest and study numerically a novel type of hybrid dimer nanoantennas. The nanoantennas consist of asymmetric metal-dielectric (Au/Si) nanoparticles and can allow tuning of the near- and far-field properties via laser melting of the metal part. We demonstrate a modification of scattering properties, near electric field distribution, normalized local density of states, and power patters of radiation of the nanoantennas upon laser reshaping. The parameters used to investigate these effects correspond to experimentally demonstrated values in the recent work.
\end{abstract}

\maketitle

The resonant plasmonic (e.g. gold, silver) nanoparticles are proven to be efficient systems for the light control at the nanoscale, due to the ability to localize optical fields via excitation of strong plasmon resonances~\cite{MaierBook, Giannini}. On the other hand, low-loss high-index dielectric (e.g. silicon, germanium) nanoparticles with inherent \textit{magnetic and electric} Mie-type resonances offer great opportunity for light controlling via designing of their scattering properties~\cite{evlyukhin2010optical, garcia2011strong, KrasnokOE, person2013demonstration, Faraon}. Recently, it has been shown, that combination of these two paradigms in form of metal-dielectric (hybrid) nanoantennas allows utilizing of the advantages of both plasmonics and all-dielectric nanophotonics~\cite{Noskov_12, Zeng2014, WangACSnano2015, Dima_AM2016, Banzer2016}. Because of the high-index dielectric nanoparticles these hybrid structures have artificial resonant magnetic response in the visible range, and the plasmonic ones provide a strong electric field enhancement. Today, the hybrid nanostructures have a plethora of applications, such as ultrahigh optical light absorption~\cite{VijayACSnano2015}, unidirectional visible light scattering~\cite{WangACSnano2015, Dima_AM2016}, plasmon-enhanced photocatalytic chemical reactions like photodegradation of organic pollutants~\cite{Jinyao_naoscale2011, Xianwei_JMCA2013}, photosynthesis of organic molecules~\cite{Atsuhiro2013, Shin_JACS2012}, enhancing and tailoring the photoluminescence~\cite{Koudo_NSRL2012, Wei_Langmuir2010, Krishna_APL2009}, surface-enhanced Raman scattering~\cite{Xiao-Dong_JRS2012, Jianhua_Langmuir2013}, and solar cells with high power conversion efficiencies~\cite{Michael_Naonolett2011, Hua2015, QQN_AM2013}.

\begin{figure*}
\includegraphics[width=0.99\textwidth]{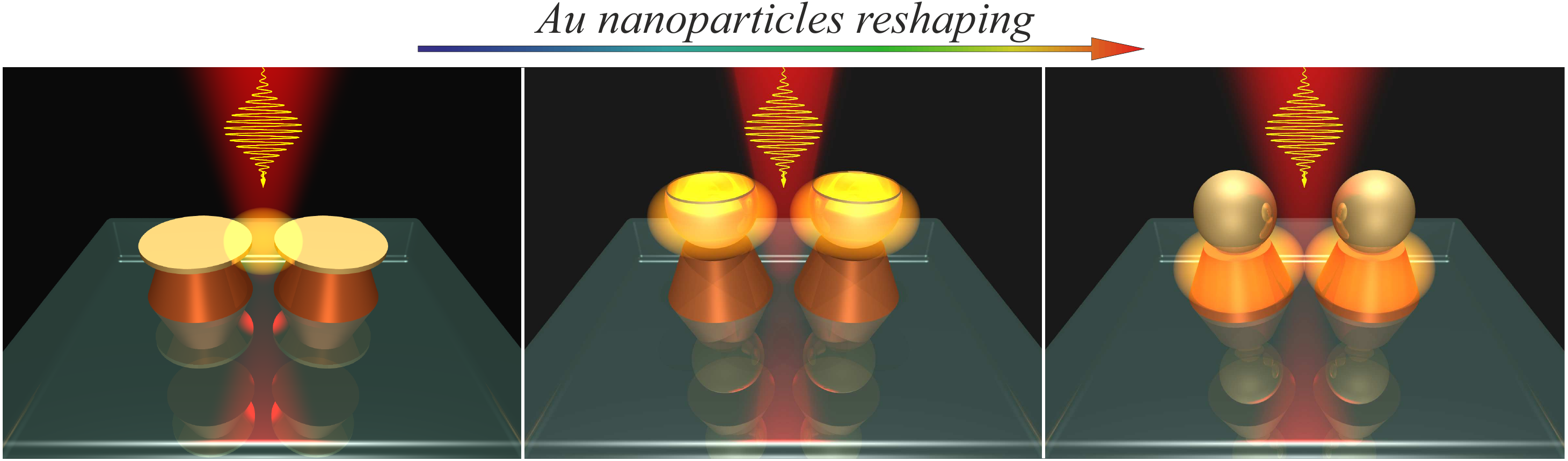}
\caption{Sketch of the dimer nanoantennas composed of hybrid Au/Si nanoparticles with different shapes of the Au nanoparticle (different levels of laser reshaping): nanodisks, nanocups, and nanospheres.}\label{artistic}
\end{figure*}

It has recently been experimentally demonstrated that asymmetric hybrid metal-dielectric (Au/Si) nanoparticles which consist of a plasmonic nanodisk placed on the top of a truncated silicon nanocone, can be tuned via fs-laser melting at the nanometer scale~\cite{Dima_AM2016}. The silicon nanocone has electric and magnetic dipole resonances, whereas the gold nanodisk has only electric one in visible range. The melting leads to the controllable reshaping of Au nanoparticle from nanodisk to nanocup, and then to nanosphere via fs-laser irradiation makes it possible to change the position of its plasmon resonance and adjust the properties of the whole hybrid nanoparticle. In this approach, it is necessary to use namely a truncated nanocone to properly change the Au nanoparticle shape. At the same time, the Si nanocone is hardly exposed to any impact of the fs-laser radiation due to the higher melting temperature and enthalpy of fusion (about 1687~K and 50.21~kJ/mol for crystalline silicon in contrast to 1337~K and 12.55~kJ/mol for gold).

A special class of nanoantennas is dimer ones, which consist of two nanoparticles divided by a subwavelength gap, and exhibit very strong enhancement of electric and magnetic fields. Such nanoantennas are relatively easy for implementation by chemical or lithography methods, and provide a possibility to engineer the local density of states in the nanogaps~\cite{Savasta2010, Bakker2015, Caldarola2015, Zywietz2015, Kuhler2014}. From the practical viewpoint, it is necessary to have an approach to tune the properties of the dimer nanoantennas. In this Article, being inspired by our recent experimental work~\cite{Dima_AM2016} on a new technique for fabrication of asymmetric hybrid nanoparticles via femtosecond laser melting at the nanoscale, we propose a novel type of hybrid dimer nanoantenna consisting of asymmetric metal-dielectric (Au/Si) nanoparticles. We suggest an approach to tune the near- and far-field properties of such nanoantennas (see Fig.~\ref{artistic}). We demonstrate numerically the modification of scattering properties, near electric field distribution, normalized local density of states, and power patters of radiation of the nanoantennas upon reshaping.

\begin{figure}[!t]
\includegraphics[width=0.5\textwidth]{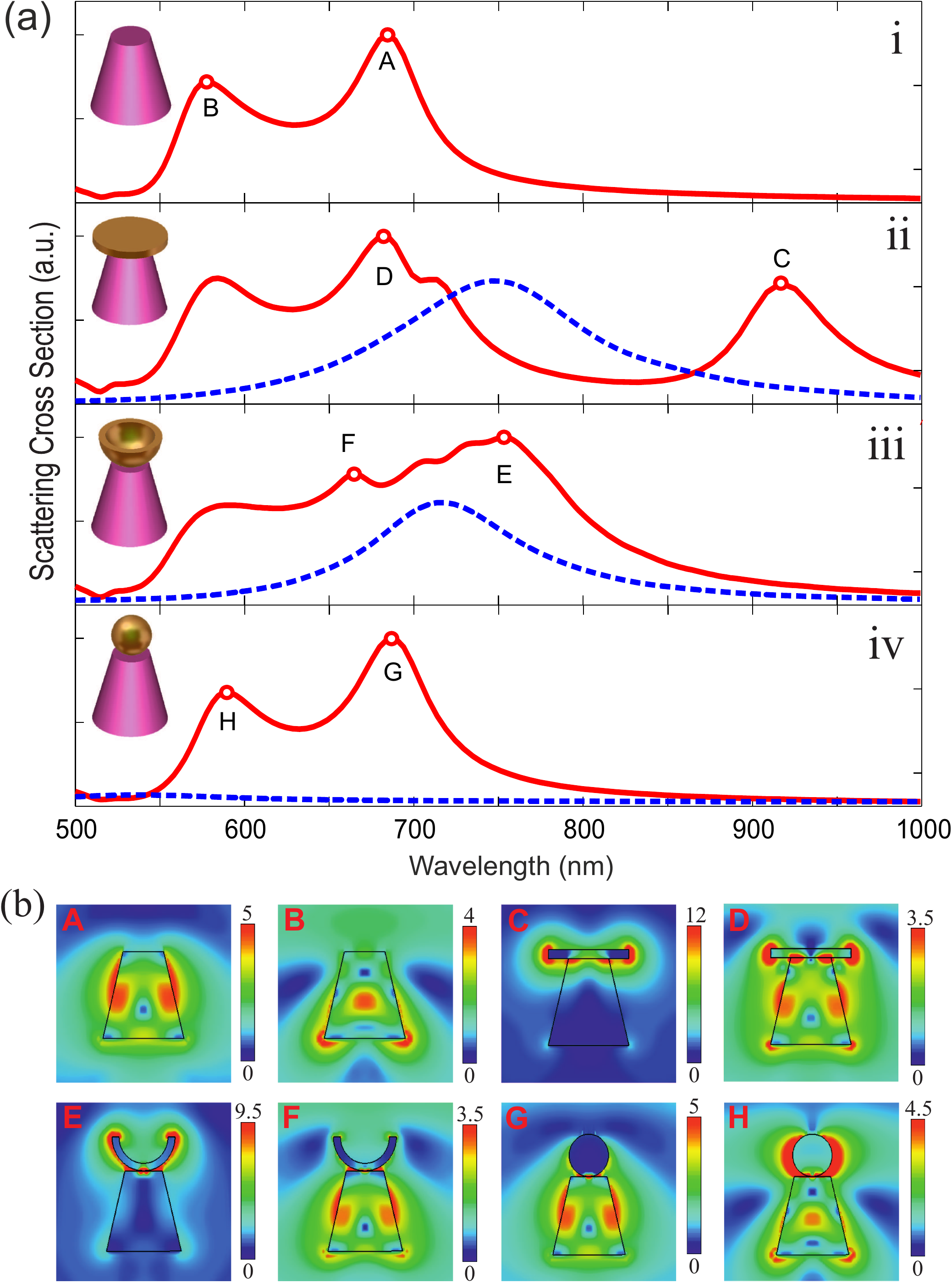}
\caption{(a)~The scattering spectra of (i)~single Si nanocone, and hybrid Au/Si nanoparticles: Si nanocone with (ii)~Au nanodisk, (iii)~Au nanocup, and (iv)~Au nanosphere, for the Si nanocone bottom diameter of 190 nm. The diameter of the Au nanodisk is equal to the diameter of the lower base of the Si nanocone. Blue curves are the scattering spectra of corresponding Au nanoparticles in free space. The incident wave propagates along the axis of symmetry of the nanoparticles. (b)~The electric field distribution for the respective cases.}\label{unit}
\end{figure}

To begin with, let us consider a scattering of a plane electromagnetic wave by a single silicon cone and hybrid nanoparticles with different level of laser reshaping (see Fig.~\ref{unit}).  We assume the wave propagation along the axis of symmetry of the nanoparticles. Fig.~\ref{unit}(a) represents the scattering spectra for various nanoparticles: (i) single Si nanocone, and hybrid Au/Si nanoparticle with (ii) Au nanodisk, (iii) Au nanocup, and (iv) Au nanosphere, in the wavelength range of 500--1000~nm. The geometric parameters of the truncated silicon cone have been taken as to be close to used in Ref.~\cite{Dima_AM2016}. The diameter of the upper base is a=95~nm, the diameter of the bottom base is 190~nm, and the height is h=190~nm. The diameter of Au nanodisk is equal to the diameter of the lower base of the Si nanocone, it follows from conditions of a lithography process~\cite{Dima_AM2016}. The thickness of Au nanodisk is $b=20$~nm. Upon the fs-laser reshaping process the gold nanoparticles keep the same volume but get the different shapes (disk, cup, and sphere) in depends on the laser intensity (see Fig.~\ref{artistic}). Therefor, since the nanoparticles volume is constant, we can conclude that the resulting nanocup has the outer and inner radii of about 80~nm and 62.2~nm, respectively. In the same manner, we infer that the resulting nanosphere has the radius about 51.3~nm.

The numerical full-wave simulations have been conducted using CST~Microwave~Studio. The dielectric permittivities of Au and Si are taken from Refs.~\cite{Christy_1972, Palik_1985, Meyer_2006}. The resonances in the scattering spectra of the single nanocone (Fig.~\ref{unit}(a)i) at points A and B are the well-known dipole electric and magnetic Mie ones inherent to dielectric nanoparticles (see the electric field distributions in Fig.~\ref{unit}(b)A,B). The magnetic dipole Mie resonance condition in the dielectric (such as silicon) nanoparticle is $\lambda_{\rm res}\approx D n_{\rm d}$, where $D$ is a characteristic size of a nanoparticle, $\lambda_{\rm res}$ is the resonant wavelength, $n_{\rm d}$ is the refractive index of the dielectric nanoparticle.

The single hybrid Au/Si nanoparticle, consisting of Si cone and Au disk, has three resonances in the scattering spectrum (Fig.~\ref{unit}(a)ii). They are the plasmon resonance of the nanodisk (Fig.~\ref{unit}(b)C) at the wavelength of 910~nm, and the dipole magnetic (Fig.~\ref{unit}(b)D) and electric Mie resonances of the Si nanocone. The plasmon resonance of the Au nanoparticle arises from excitation of the localized surface plasmon mode, and it strongly depends on its geometric parameters and shape. Recently, it has been shown that under irradiation of a hybrid nanoparticle by femtosecond laser pulse with energy of $40-50$~mJ/cm$^2$, the Au nanoparticle changes its form from a nanodisk to a nanocup~\cite{Dima_AM2016}. At intensities below these values, the particle scatters light without changing its shape. The plasmon resonance of the resulting nanoparticle [see scattering spectra in Fig.~\ref{unit}(a)iii, and the electric field distribution in Fig.~\ref{unit}(b)E] shifts to shorter wavelengths (from 910~nm to 750~nm, in our case). By increasing the energy density of the laser radiation up to $70$~mJ/cm$^2$, the nanocup takes the shape of a nanosphere. In this case, the plasmon resonance shifts to more shorter wavelengths (590~nm), where it is approximately equals the electric dipole Mie resonance of the Si nanocone [see scattering spectra in Fig.~\ref{unit}(a)iv, and the electric field distribution in Fig.~\ref{unit}(b)H]. Thus, the position of the hybrid nanoparticle plasmon resonance can be precisely controlled via fs-laser reshaping. Blur curves in Fig.~\ref{unit} indicates the scattering spectra of corresponding Au nanoparticles arranged in free space. Being placed on the high-index nanocone the plasmon resonances of Au nanoparticles shift to the longer wavelengths. It is seen, that the resonant wavelength of the Au particle shifts upon laser reshaping with a strong decrease in the scattering amplitude. For example, Au nanosphere have a weak effect on scattering of the entire hybrid nanoparticle, whereas it affect strongly on the near field distribution (compare Fig.~\ref{unit}(b)B and Fig.~\ref{unit}(b)H).

\begin{figure}[!t]
\includegraphics[width=0.5\textwidth]{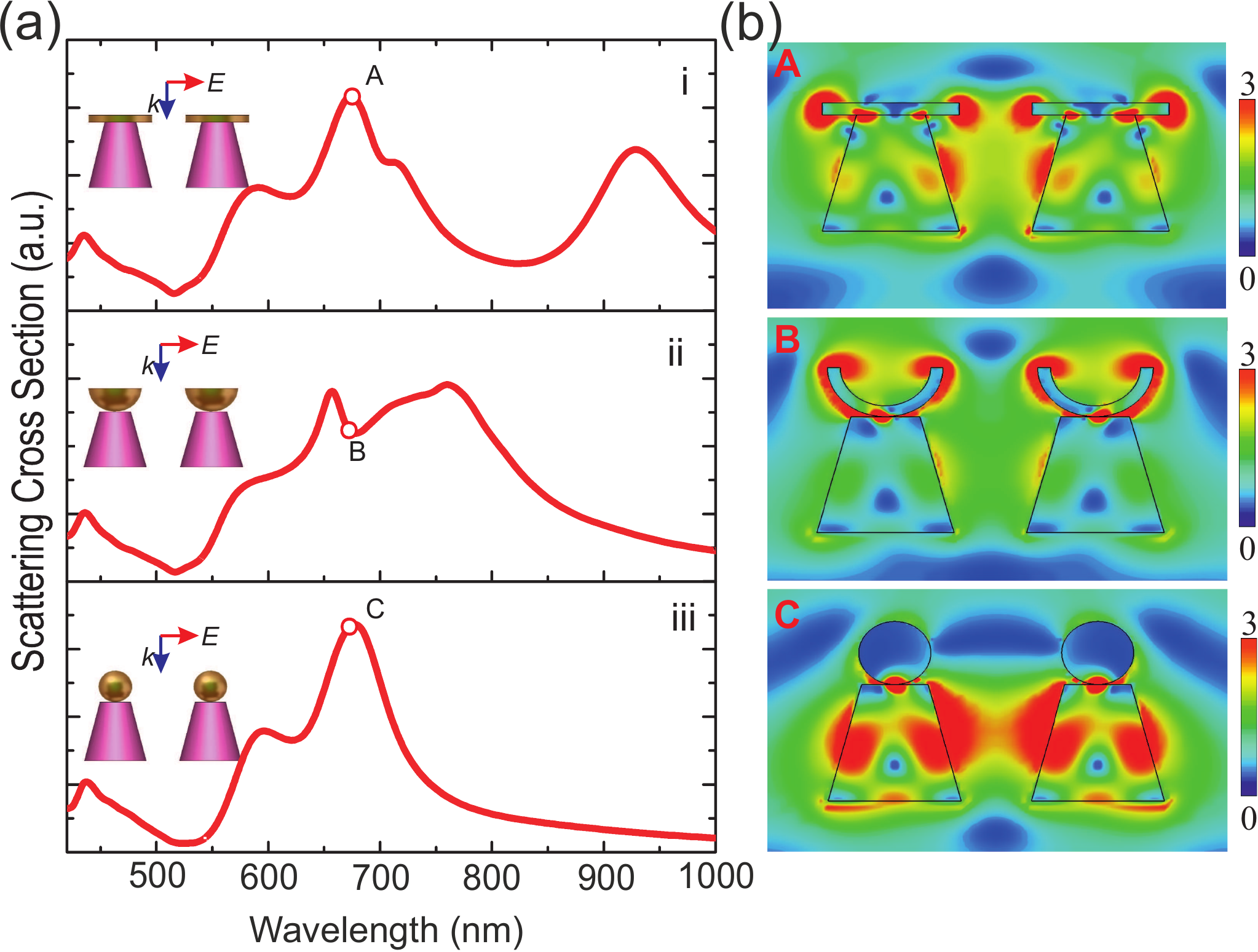}
\caption{(a)~The scattering spectra of hybrid dimer nanoantennas with different degrees of the fs-laser reshaping of Au nanoparticles: (i)~nanodisks, (ii)~nanocups, and (iii)~nanospheres. (b)~The distributions of the local E-field for wavelength of 670~nm. The incident wave propagates as shown in the insets.}
\label{RCS}
\end{figure}

Now, let us show that the laser reshaping strongly influences the scattering properties of the dimer nanoantennas consisting of two hybrid Au/Si nanoparticles divided by a subwavelength gap (as it schematically shown in Fig.~\ref{artistic}). We consider scattering of a plane electromagnetic wave by the hybrid Au/Si dimer nanoantennas (see Fig.~\ref{RCS}). The incident wave propagates as shown in the insets. Fig.~\ref{RCS}(a) shows the scattering spectra of the hybrid dimer nanoantennas with different degrees of the fs-laser reshaping of Au nanoparticles: (i) -  nanodisks, (ii) - nanocups, and (iii) - nanospheres, in the range of 420--1000~nm. The geometrical parameters of the hybrid nanoparticles are the same as in Fig.~\ref{unit}. The distance between the axes of the hybrid nanoparticles ($D$) is 100~nm in all cases. Fig.~\ref{RCS}(a)i shows the scattering spectrum of the hybrid dimer nanoantanna without reshaping (the gold nanoparticle has a disk shape). The next plot (Fig.~\ref{RCS}(a)ii) is the scattering spectrum of the nanoantenna with gold particles in the form of nanocup. The third plot (Fig.~\ref{RCS}(a)iii) is the scattering spectrum of the nanoantenna in the final stage of reshaping. It is seen that the scattering spectrum dramatically changes with reshaping of the Au nanoparticles in the hybrid nanoantenna. Upon reshaping, the electric dipole resonance of the Au nanoparticle shifts to magnetic dipole resonance (point A in Fig.~\ref{RCS}(a)i) of silicon one. The scattering spectrum of the hybrid dimer nanoantenna with gold nanocups (see Fig.~\ref{RCS}(a)ii) is more complicated because of a complex shape of the Au nanoparticles. In Fig.~\ref{RCS}(b) the electric field distributions for the hybrid dimer nanoantennas with varying degrees of reshaping are presented. We observe an increasing of a local electric field in the gap of the hybrid nanoantanna by 3 times. Moreover, the electric field distribution strongly depends on the Au nanoparticles shape. For example, the difference in E-field straight in the middle of the nanoantenna reaches 2, for the cases of nanocups and nanospheres. Thus, we have numerically shown that the laser melting of the hybrid nanoantenna makes it possible to tune its far-field (scattering) and near-field (electric field distribution) properties.

\begin{figure}[!t]
\includegraphics[width=0.5\textwidth]{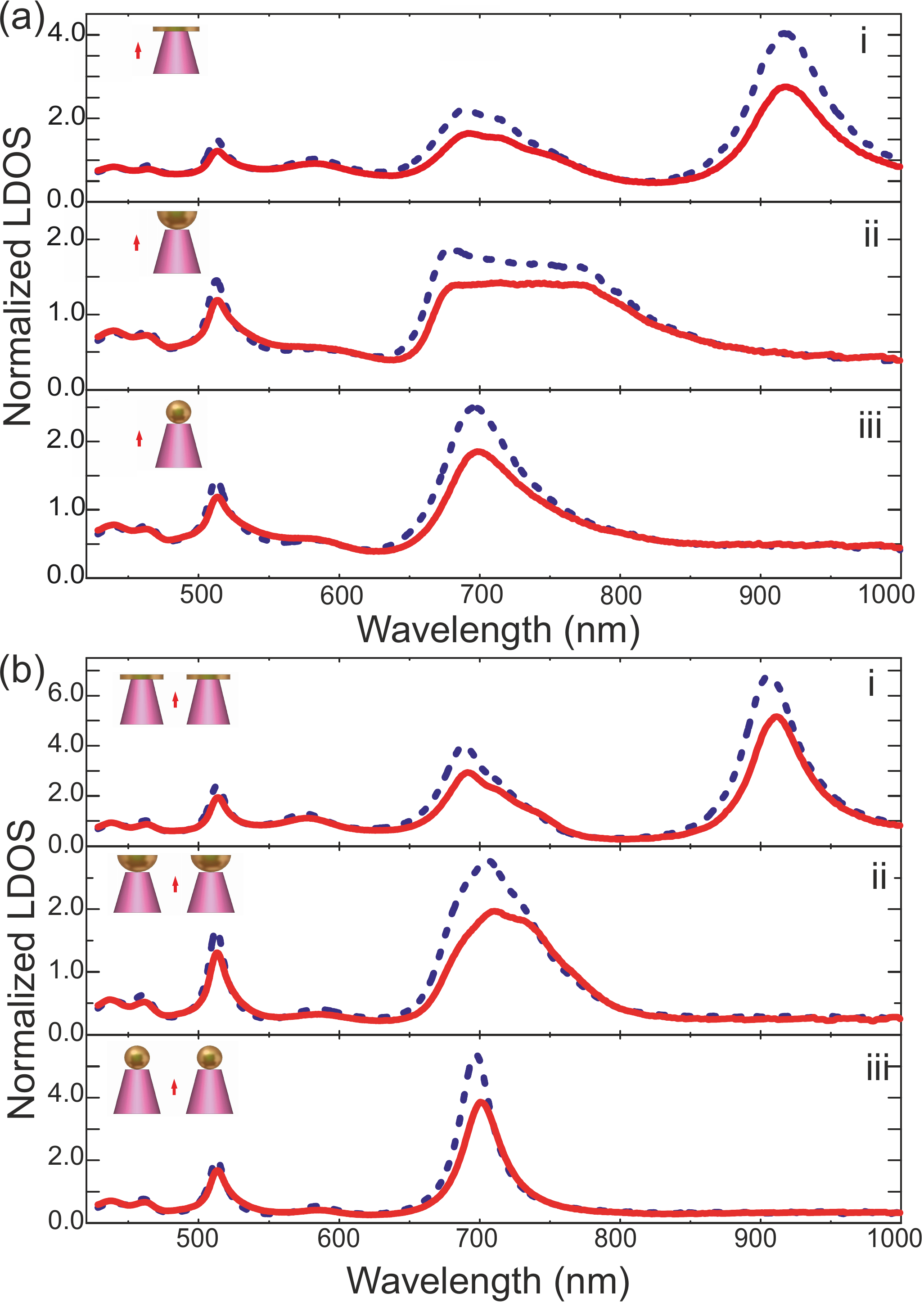}
\caption{The normalized local density of states of hybrid nanoantennas with different fs-laser reshaping degree as a function of operation wavelength: (a)~single nanoparticles, (b)~hybrid dimer nanoantennas; for different fs-laser reshaping degree: (i)~Au nanodisks, (ii)~Au nanocups, and (iii)~Au nanospheres. The blue lines correspond to $G=125$~nm and the red lines correspond to $G=145$~nm, where $G$ is the distance between the dipole source and the hybrid nanoparticle axis. The dipole source position and orientation depicted in the insets.}
\label{purcell}
\end{figure}

Now, we turn to show that asymmetric hybrid nanoparticles and dimer nanoantennas have ability to tune the \textit{Purcell effect} via laser melting of the Au components. The Purcell effect is manifested in a modification of the spontaneous emission rate of a quantum emitter induced by its interaction with inhomogeneous environment (dimer nanoantenna, in our case) and is expressed by the Purcell factor $F$ (or \textit{normalized local density of states})~\cite{Purcell_46, Sauvan2013a, BarthesPhysRevB2011, Poddubny2012, Krasnok_Purcell_2015, KrasnokAPL2016}. This modification is significant if the environment is a resonator tuned to the emission frequency. Although the Purcell effect was discovered in the context of nuclear magnetic resonance~\cite{Purcell_46}, nowadays it is widely used in many applications, ranging from microcavity light-emitting devices~\cite{Fainman_2010} to single-molecule optical microscopy~\cite{Koenderink_PRL_11, Cosa_2013}, being also employed for tailoring optical nonlinearities~\cite{Soljacic_2007} and enhancing spontaneous emission from quantum cascades~\cite{Minot_2007}. Open nanoscale resonators such as plasmonic nanoantennas can change the spontaneous emission lifetime of a single quantum emitter, that is very useful in microscopy of single NV centers in nanodiamonds~\cite{Vamivakas_NanoLetters_13}, Eu$^{3+}$-doped nanocrystals~\cite{Carminati_PRL_14}, plasmon-enhanced optical sensing~\cite{Sauvan2013a}, and the visualization of biological processes with large molecules~\cite{Tinnefeld_Science_2012}.

The normalized local density of states (LDOS) is usually determined by the value of imaginary part of the Green's function of a point source located at the point $\mathbf{r}_0$ on itself~\cite{KrasnokAPL2016}, i.e., by the value ${\rm Im}\left[G_{zz}(r_0,r_0,\omega)\right]$, where $\omega$ is radiation frequency, and we assume the source oriented along the axis $z$. Our method of calculation of the normalized LDOS of such complex systems in \textit{the stationary regime} is based on the relationship between the input impedance of a small (in terms of the radiation wavelength) dipole emitter and its radiated power, which is completely equivalent to the Green's function method~\cite{SlobozhanyukAPL2014, Krasnok_Purcell_2015}:
\begin{equation}
F=\frac{{\rm Im}\left[G_{zz}(r_0,r_0,\omega)\right]}{{\rm Im}\left[G^{0}_{zz}(r_0,r_0,\omega)\right]}=\frac{{\rm Re} [Z_{{\rm in}}]}{{\rm Re} [Z_{0, {\rm in}}]}.
\label{eq:1}
\end{equation}
Here, ${\rm Re} [Z_{{\rm in}}]$ and ${\rm Re} [Z_{0, {\rm in}}]$ are the real part of the input impedance of the small dipole source in the presence of the hybrid nanoantenna and in free space, respectively. The value $G^{0}_{zz}$ is the Green's function of free space referred to the same point. Obviously, ${\rm Re} [Z_{0, {\rm in}}]$ is a constant (at a certain wavelength) for a specific dipole, and then to improve the Purcell factor, we just need to concentrate on ${\rm Re} [Z_{{\rm in}}]$ according to~\eqref{eq:1}. Fig.~\ref{purcell}(a) represents the calculated normalized LDOS of a single hybrid nanoparticle in the wavelength range of 420-1000~nm. The dipole source position and orientation depicted in the insets. The dipole emitter (red arrow) is modeled as a Hertzian dipole of length 20~nm which is less than $\lambda $/15. The dipole source is located at the distance $G$ from the hybrid nanoparticle axis. We observe the sharp effect of the Au nanoparticle reshaping on the normalized LDOS of the hybrid monopole nanoantenna. We also demonstrate the normalized LDOS enhancement with reducing of the distance $G$.

\begin{figure}[!t]
\includegraphics[width=0.5\textwidth]{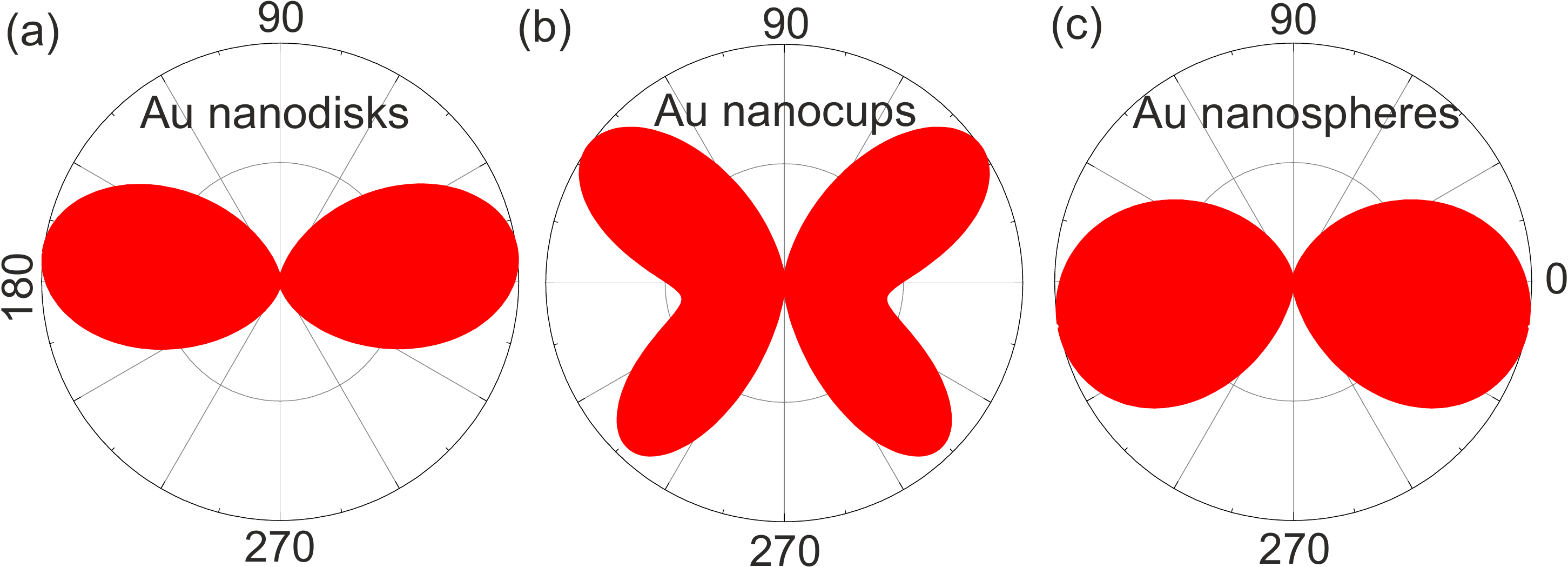}
\caption{The power patterns of radiation of the dipole source placed in the middle of the gap of the hybrid (Au/Si) dimer nanoantennas with different shapes of the Au component: (a)~nanodisks, (b)~nanocups, and (c)~nanospheres. The wavelength of radiation is 750~nm.}
\label{patterns}
\end{figure}

Fig.~\ref{purcell}(b) shows the calculated normalized LDOS of the proposed hybrid dimer nanoantennas in the same spectral range. By comparing Figs.~\ref{purcell}(a) and (b), we observe the growth of the normalized LDOS in the case of the dimer nanoantenna compared with the monomer. The value of the normalized LDOS strongly depends on the gold nanoparticle shape. Moreover, we observe the dependence of the spectral width of the resonances on the shape of the Au particles. For example, the spectral width of the magnetic dipole resonance of the dimer nanoantenna with Au nanocups is 100~nm (Fig.~\ref{purcell}(b)ii), whereas in the case of Au nanospheres it is equal 50~nm (Fig.~\ref{purcell}(b)iii). In results in enhancement of the Purcell factor by 2 times upon laser reshaping. The distance $G$ is a critical factor here. Lower distance, namely the hybrid nanoantenna is more adjacent to the dipole, brings in stronger oscillation, and thus results in higher Purcell factor as clearly indicated in Fig.~\ref{purcell}. Apparently, the hybrid dimer nanoantennas can result in more intense interactions between two components, and thus achieve higher Purcell factor which opens up a new way to future nanoscale optical designs.

Finally, Fig.~\ref{patterns} shows the results of calculation of the power patterns for the dipole source placed in the middle of the gap of the hybrid dimer nanoantennas with different shapes of the Au component: nanodisks, nanocups, and nanospheres (Figs.~\ref{patterns}(a)--(c), respectively). The wavelength of dipole radiation is 750~nm. We observe the strong influence of the laser reshaping on the radiation patterns of the dipole source placed in the vicinity of the hybrid dimer nanoantennas. The nanoantenna exhibits such strong dependence of the power patterns at the other wavelengths in the vicinity of resonances.

In summary, we have proposed and studied numerically a novel type of hybrid dimer nanoantennas consisting of asymmetric metal-dielectric (Au/Si) nanoparticles that exhibit magnetic and electric dipole resonances in the visible range. We have proposed a practical approach to tune the near- and far-field properties of such nanoantennas via fs-laser melting at the nanoscale. We have demonstrated numerically the modification of scattering properties, near electric field distribution, normalized local density of states, and power patters of radiation of the nanoantennas upon reshaping. We believe that our results provide an attractive platform for various nanophotonics applications.

This work was financially supported by Russian Science Foundation (Grant 15-19-30023). The authors are thankful to Mr.~Mingzhao Song, Dr.~Denis Baranov, and Dr.~Andrey Miroshnichenko for useful discussions.


%

\end{document}